\documentclass[final,3p,times,twocolumn]{elsarticle}

\usepackage{amssymb}
\usepackage{hyperref}

\journal{Astroparticle Physics}

\begin{document}

\begin{frontmatter}

\title{A search for solar axion induced signals with COSINE-100}

\author[addr1]{P.~Adhikari\fnref{fn1}}
\author[addr1]{G.~Adhikari}
\author[addr2]{E.~Barbosa de Souza}
\author[addr3]{N.~Carlin}
\author[addr4]{S.~Choi}
\author[addr5]{M.~Djamal}
\author[addr6]{A.C.~Ezeribe}
\author[addr7]{C.~Ha}
\author[addr8]{I.S.~Hahn}
\author[addr7]{E.J.~Jeon}
 \author[addr2]{J.H.~Jo}
 \author[addr4]{H.W.~Joo}
 \author[addr7]{W.G.~Kang}
 \author[addr9]{W.~Kang}
 \author[addr10]{M.~Kauer}
 \author[addr11]{G.S.~Kim}
 \author[addr7]{H.~Kim}
 \author[addr11]{H.J.~Kim}
 \author[addr7]{K.W.~Kim}
 \author[addr7]{N.Y.~Kim}
 \author[addr4]{S.K.~Kim}
 \author[addr7,addr1,addr12]{Y.D.~Kim}
 \author[addr7,addr13,addr12]{Y.H.~Kim}
 \author[addr7]{Y.J.~Ko}
 \author[addr6]{V.A.~Kudryavtsev}
 \author[addr7,addr12]{H.S.~Lee}
 \ead{hyunsulee@ibs.re.kr}
 \author[addr7]{J.~Lee}
 \author[addr11]{J.Y.~Lee}
 \author[addr7,addr12]{M.H.~Lee}
 \author[addr7]{D.S.~Leonard}
 \author[addr6]{W.A.~Lynch}
 \author[addr2]{R.H.~Maruyama}
 \author[addr6]{F.~Mouton}
 \author[addr7]{S.L.~Olsen}
 \author[addr12]{B.J.~Park}
 \author[addr14]{H.K.~Park}
 \author[addr13]{H.S.~Park}
 \author[addr7]{K.S.~Park}
 \author[addr3]{R.L.C.~Pitta}
 \author[addr5]{H.~Prihtiadi}
 \author[addr7]{S.~Ra}
 \author[addr9]{C.~Rott}
 \author[addr7]{K.A.~Shin}
 \author[addr6]{A.~Scarff}
 \author[addr6]{N.J.C.~Spooner}
 \author[addr2]{W.G.~Thompson}
 \author[addr15]{L.~Yang}
 \author[addr9]{G.H.~Yu}
 \author[]{(COSINE-100 Collaboration)}
 \address[addr1]{Department of Physics and Astronomy, Sejong University, Seoul 05006, Korea}
 \address[addr2]{Wright Laboratory, Department of Physics, Yale University, New Haven, CT 06520, USA} 
 \address[addr3]{Physics Institute, University of S\~{a}o Paulo, 05508-090, S\~{a}o Paulo, Brazil}
 \address[addr4]{Department of Physics and Astronomy, Seoul National University, Seoul 08826, Republic of Korea}
 \address[addr5]{Department of Physics, Bandung Institute of Technology, Bandung 40132, Indonesia}
 \address[addr6]{Department of Physics and Astronomy, University of Sheffield, Sheffield S3 7RH, United Kingdom}
 \address[addr7]{Center for Underground Physics, Institute for Basic Science (IBS), Daejeon 34126, Republic of Korea}
 \address[addr8]{Department of Science Education, Ewha Womans University, Seoul 03760, Republic of Korea}
 \address[addr9]{Department of Physics, Sungkyunkwan University, Seoul 16419, Republic of Korea}
 \address[addr10]{Department of Physics and Wisconsin IceCube Particle Astrophysics Center, University of Wisconsin-Madison, Madison, WI 53706, USA}
 \address[addr11]{Department of Physics, Kyungpook National University, Daegu 41566, Republic of Korea}
 \address[addr12]{IBS School, University of Science and Technology (UST), Daejeon 34113, Republic of Korea}
 \address[addr13]{Korea Research Institute of Standards and Science, Daejeon 34113, Republic of Korea}
 \address[addr14]{Department of Accelerator Science, Graduate School, Korea University, Sejong 30019, Korea}
 \address[addr15]{Department of Physics, University of Illinois at Urbana-Champaign, Urbana, IL 61801, USA}

 \fntext[fn1]{Present address: Department of Physics, Carleton University, Ottawa, Ontario, K1S 5B6, Canada}

\begin{abstract}
		We present results from a search for solar axions with the COSINE-100 experiment. 
		We find no evidence of solar axion events from a data-set of 6,303.9 kg$\cdot$days exposure and set a 90\,\% confidence level upper limit on the axion-electron coupling, $g_{ae}$, of 1.70~$\times$~$10^{-11}$ for an axion mass less than 1\,keV/c$^2$. This limit excludes QCD axions heavier than 0.59\,eV/c$^2$ in the DFSZ model and 168.1\,eV/c$^2$ in the KSVZ model.
\end{abstract}


\end{frontmatter}

\section{Introduction}
\label{intro}
The axion, a pseudo-Nambu-Goldstone boson introduced by Wilczek~\cite{FWilczeck} and Weinberg~\cite{SWeinberg}, appears in the Peccei-Quinn solution of the strong CP problem~\cite{RDPeccei}. Even though the original axion model~\cite{FWilczeck,SWeinberg} was ruled out by laboratory experiments, KSVZ (Kim-Shifman-Vainstein-Zakharov)~\cite{JEKim70,Shifman} and DFSZ (Dine-Fischler-Srednicki-Zhitnitskii)~\cite{ARZh,Dine} {\it invisible} axion models are not excluded by either terrestrial experiments or astrophysics considerations~\cite{pdg}. 

\begin{figure*}[!htb]
\centering
\begin{tabular}{cc}
\includegraphics[width=0.42\textwidth]{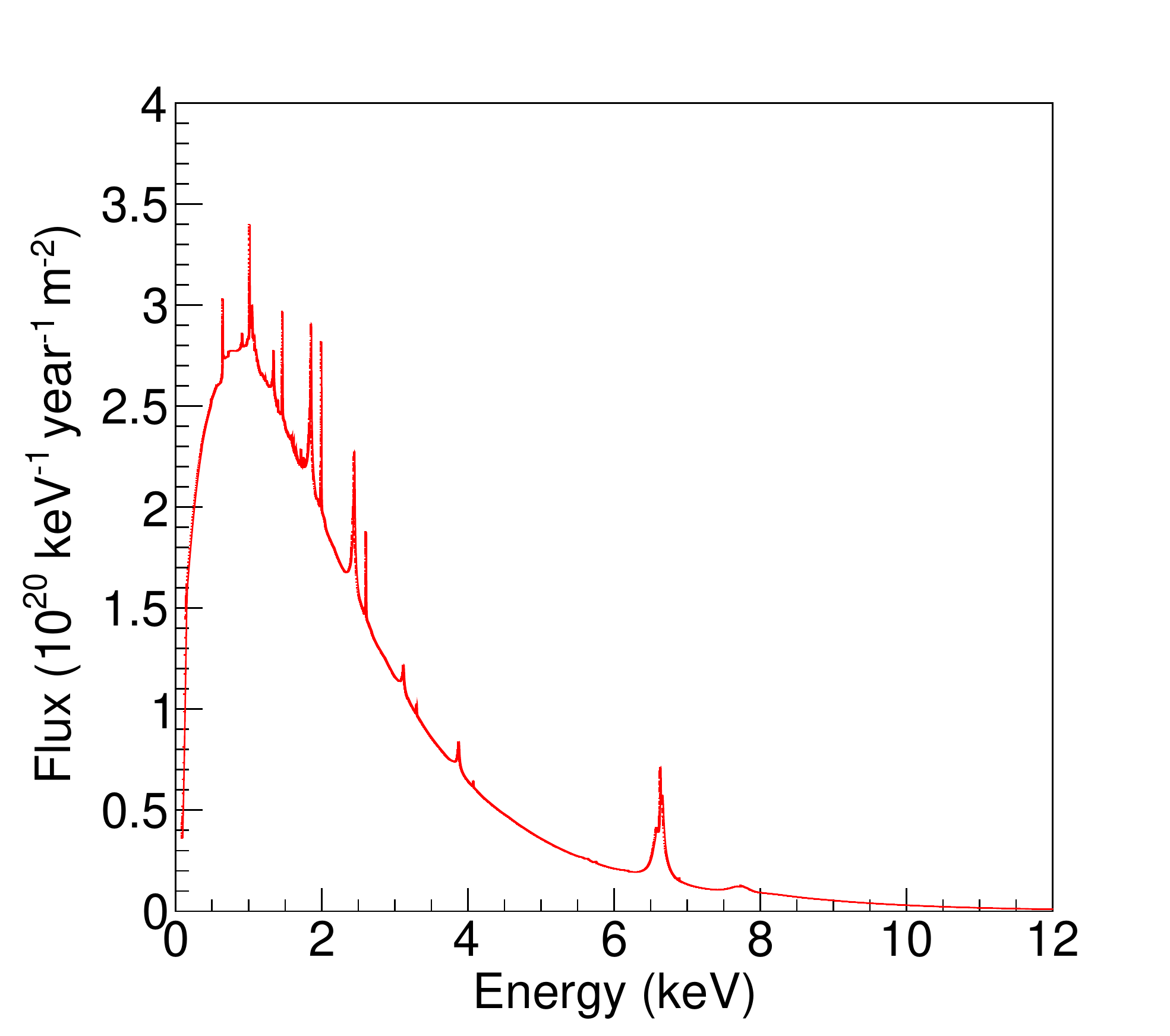} &
\includegraphics[width=0.54\textwidth]{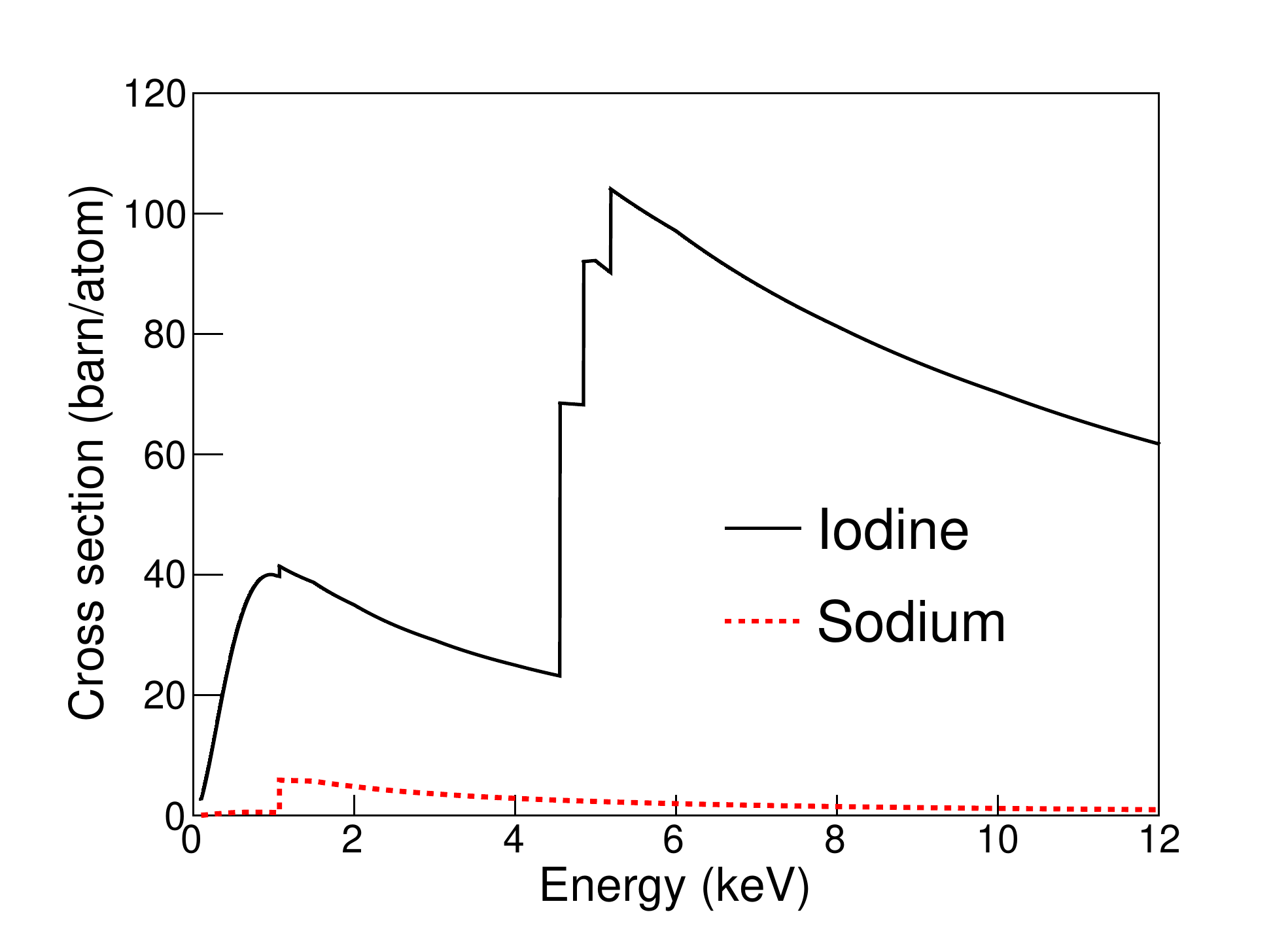} \\
      (a) & (b)\\
\end{tabular}
 \caption{ (Color online) (a) The flux of solar axions on the Earth considering the Compton scattering, axio-recombination, axio-deexcitation, axio-bremsstrahlung and electron-electron collision mechanisms~\cite{JRedondo013} for an axion-electron coupling~($g_{ae}$) value of $10^{-13}$ is shown. (b) Axio-electric cross sections for Na (dotted red line ) and I (solid black line) atoms are presented for $m_a$ = 0.0\,keV/c$^2$ and $g_{ae}$  = 1.}
\label{Axion_Flux_earth}
\end{figure*}

In these models, astrophysical objects such as the Sun would be intense sources of axions~\cite{PhysRevLett.51.1415} that would be produced via the following processes~\cite{JRedondo013}:
\begin{itemize}
\item Compton scattering: $\gamma + e \rightarrow e + a$
\item Axio-recombination: $e + A  \rightarrow  A^- + a$
\item Axio-deexcitation:  $A^\star \rightarrow  A + a$
\item Axio-bremsstrahlung: $e + A \rightarrow  e + A + a$
\item Electron-electron collision: $e + e \rightarrow  e + e + a$
\end{itemize}
where $e$ is an electron, $a$ is an axion and $A$ is an atom. 
The total solar axion flux on the surface of the Earth is estimated in Ref.~\cite{JRedondo013} and shown in Fig.~\ref{Axion_Flux_earth} (a).

In this paper, we present results from a solar axion search performed with the COSINE-100 experiment~\cite{COSINE100:First,COSINE100:Nature,COSINE100:Modulation}. 
A data-set corresponding to a 59.5\,day exposure with a 106\,kg array of low background NaI(Tl) crystals is analyzed.
We assume that axions are produced in the Sun and propagate to the Earth. We restrict the search to axion masses~($m_a$) that are less than 1\,keV/c$^2$ to match the validity range of the flux calculations~\cite{JRedondo013}. 

Axions could produce signals in the NaI(Tl) crystals through their coupling to photons~($g_{a\gamma}$), electrons~($g_{ae}$), and nuclei~($g_{aN}$).
The coupling $g_{ae}$ can be observed via scattering off atomic electrons in the crystals by the axioelectric effect~\cite{LMKrauss84,SDimopoulos86,PhysRevD.35.2752,Pospelov,ADerevianko10}, 
$a + A \rightarrow e^− + A^+$
where $A$ is either a Na or I atom. 
This favors DFSZ~(non-hadronic) axions that have direct couplings to leptons; KSVZ~(hadronic) axions do not have tree-level couplings to leptons and are strongly suppressed~\cite{Derbin:2012yk}. 
The cross section for axio-electric interactions~\cite{Pospelov,ADerevianko10} is,
\begin{equation} \label{axioelectric_CR}
 \sigma_{ae}(E_a)= \sigma_{pe}(E_a) \frac{g_{ae}^2}{\beta_a} \frac{3E_a^2}{16 \pi \alpha m_e^2} (1 - \frac{\beta_a^{2/3}}{3}),
\end{equation}
where $E_a$ is the axion energy, $\sigma_{pe}$ is the photoelectric cross section in either Na or I~\cite{PhysRefData}, $g_{ae}$ is the axion-electron coupling, $\beta_a$ is the axion velocity relative to the speed of light, $\alpha$ is the fine structure constant, and $m_e$ is the electron mass.
Cross sections for the case of $g_{ae}$ = 1 for both Na and I atoms are shown in Fig.~\ref{Axion_Flux_earth} (b). 

\section{COSINE-100 setup and data analysis}
\subsection{COSINE-100 experiment}
\begin{figure}[!htb]
\begin{center}
\includegraphics[width=0.9\columnwidth]{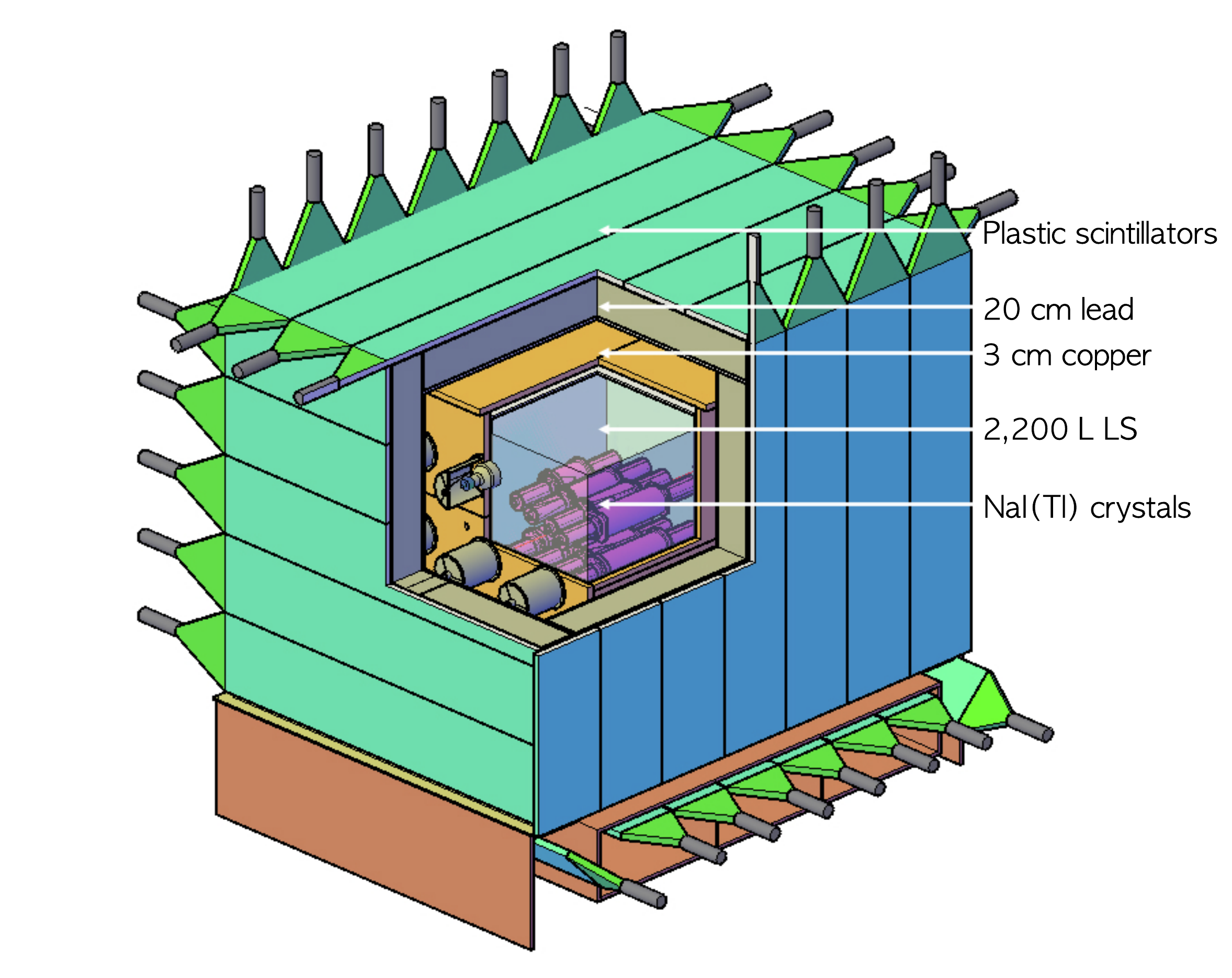}
\caption{
		Schematic of the COSINE-100 experiment. The NaI(Tl)~(106~kg) detectors are immersed in a 2,200\,L LAB-LS bath that is surrounded by  successive layers of shielding.
}
\label{fig_det}
\end{center}
\end{figure}

COSINE-100 is a dedicated experiment to test the observation of annual modulation in the event rate observed by DAMA/LIBRA experiment~\cite{dama1,dama2} utilizing  a 4~$\times$~2 array of NaI(Tl) crystals with a total mass of 106\,kg located at the Yangyang underground laboratory. The crystals are immersed in 2,200\,L of linear alkyl benzene~(LAB)-based liquid scintillator~(LS), which acts as a veto for multiple-hit events~\cite{jspark}. Shielding structures of copper, lead, and plastic scintillator surround the liquid scintillator to reduce the background contribution from external radiation and to veto cosmic-rays~\cite{COSINE100:First, COSINE:Muon} as shown in Fig.~\ref{fig_det}. Data from the two-month period between 20 October 2016 and 19 December 2016 are used in this analysis. During the period, the detector operation was very stable and there were no environmental abnormalities. 

Each crystal is coupled to two high quantum efficiency photomultiplier tubes~(PMTs), model R12669SEL from Hamamatsu photonics, that are specially selected for their high quantum efficiency. An event is triggered when coincident single photoelectrons in both PMTs from a single crystal are observed within a 200\,ns time window. If an event is triggered, waveforms from all eight crystals are recorded during 8\,$\mu$s windows around the hit~\cite{Adhikari:2018fpo}. The eight crystals are named crystal-1 to crystal-8. 
Six crystals have light yields of approximately 15\,photoelectrons/keV so that 2\,keV analysis thresholds can be applied. However, two crystals, crystal-5 and crystal-8, have lower light yields and present 4\,keV and 8\,keV analysis thresholds, respectively~\cite{COSINE100:First,COSINE100:Nature}. 
The energy calibration was done with two internal contamination in the crystals: the 3.2\,keV X-rays from $^{40}$K and the 46.5\,keV $\gamma$-rays from $^{210}$Pb.

\subsection{Event selection}
Because of the low cross-section for solar axion interactions, a solar axion may interact at most only once while traversing the COSINE-100 detector. 
An axion candidate event is defined as a signal in one crystal and no signal in any other crystals, the liquid scintillator or the muon detector. 
Events with hits in other crystals or the liquid scintillator are selected as multiple-hit events and 
used for the development of event selection criteria, energy calibration, efficiency determination and background assessment as described in Refs.~\cite{COSINE100:Nature,cosinebg, adhikarip}.

\begin{figure*}[tbp]
\centering
\begin{tabular}{ccc}
\includegraphics[width=0.33\textwidth]{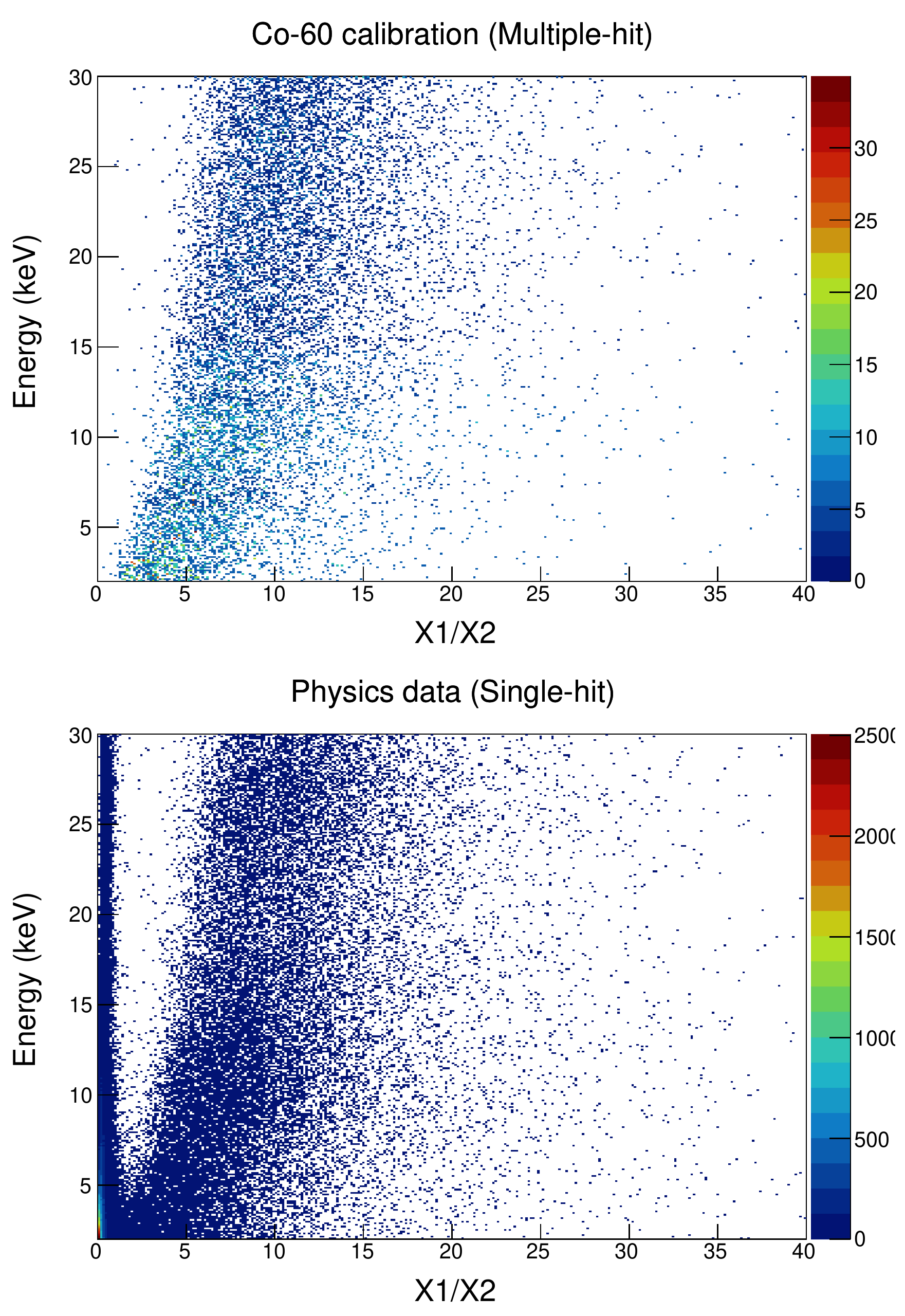} &
\includegraphics[width=0.33\textwidth]{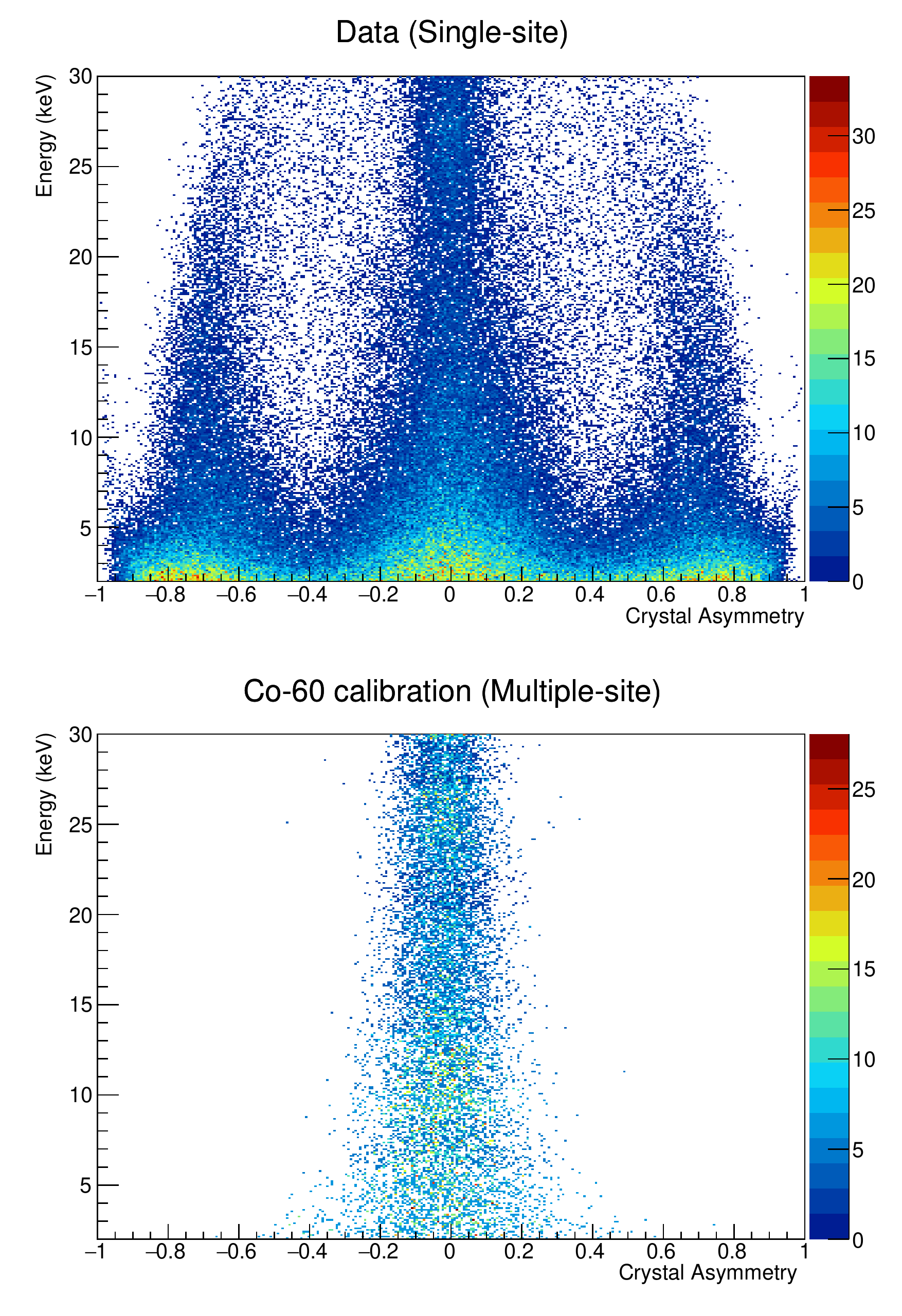} &
\includegraphics[width=0.33\textwidth]{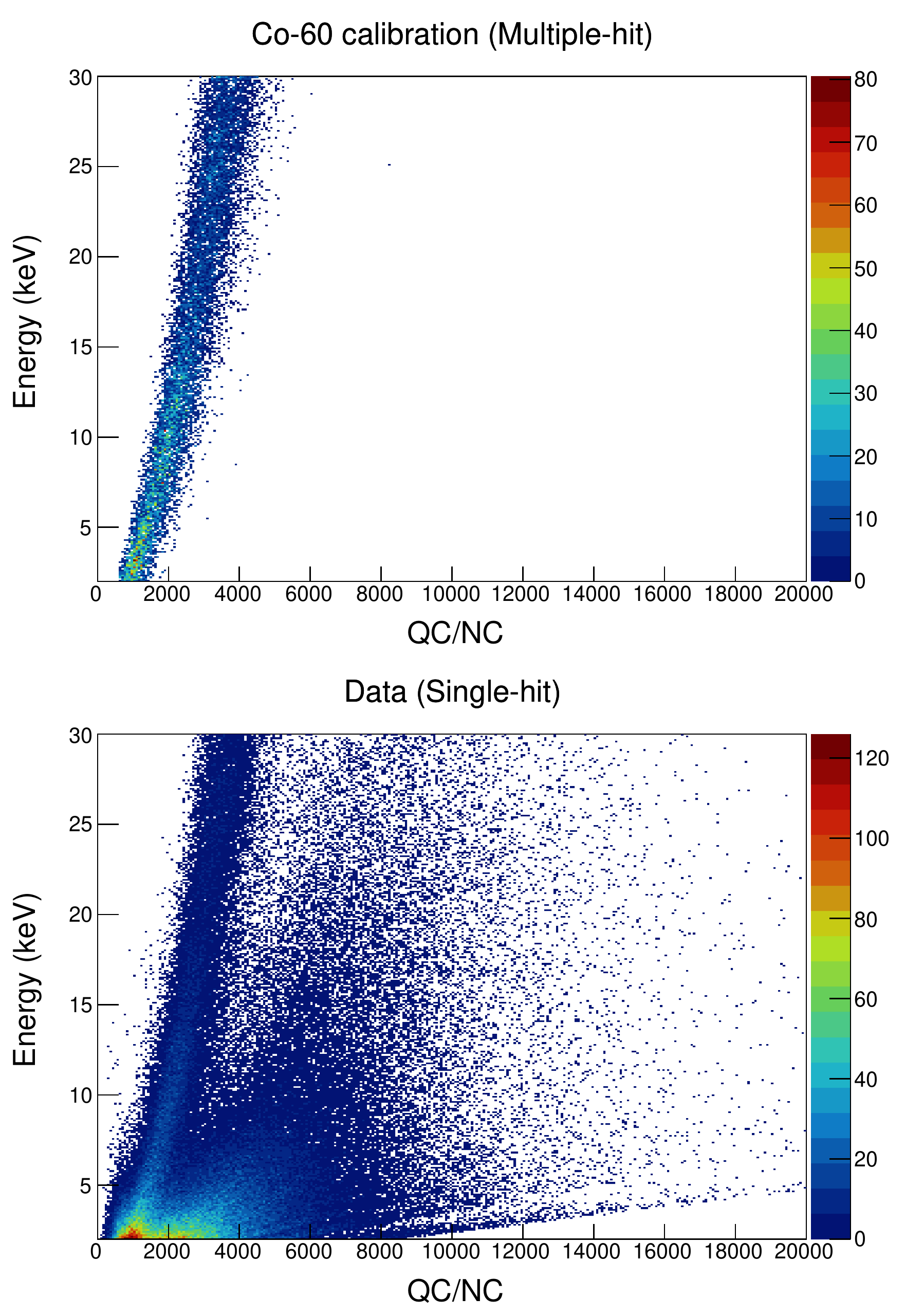} \\
(a) X1/X2  & (b) Asymmetry  & (c) QC/NC \\
\end{tabular}
\caption{\label{fig_paras}
Event selection parameters: the ratio of fast and slow charge~(a); the balance of accumulated charge between two PMTs (b); and the average charge of each cluster~(c); developed to remove PMT-induced noise are presented for the scintillation dominant $^{60}$Co calibration data~(top) and the PMT noise dominant single-hit physics data (bottom). 
}
\end{figure*}

The low-energy single-hit events corresponding to the solar axion search data-set are mostly non-physics events that are primarily caused by PMT-induced noise that is coincident between two PMTs coupled to different ends of the same crystal. 
These coincident noises could be due to radioactivity in the PMT glass and/or circuitry, the discharge of accumulated space charge in the PMTs, PMT dark current, and after-pulses from gas ionization inside the PMT~\cite{COSINE100:First,Kim:2014toa}. 
Fortunately, pulse shapes of the PMT-induced noise are significantly different from the NaI(Tl) scintillation signals. 
The DAMA group developed an efficient cut based on the ratio between ``fast'' charge (0-50\,ns), X1, and ``slow'' charge (100-600\,ns), X2~\cite{Bernabei:2008yh,Bernabei:2012zzb}. 
We also demonstrate that the ratio of X1 and X2~(X1/X2) is an essential parameter to remove the PMT noise events~\cite{COSINE100:First,Kim:2014toa}. 
However, only applying X1/X2 selection is not enough to remove all of the noise events. For this, we needed to require a balance of the deposited charge from  two PMTs~(Asymmetry) and a limit on mean charge of each cluster~(QC/NC) for efficient selection of the scintillating signals~\cite{COSINE100:First,Kim:2014toa}. 
Comparisons of the three parameters, discussed above, between radiation induced scintillation dominant  $^{60}$Co calibration data and the PMT induced noise dominant single-hit physics data  are shown in Fig.~\ref{fig_paras}. (The single-hit physics data also contain the scintillating signals as well.) 
Each parameter has a good ability to discriminate between the PMT noise pulses and the scintillating-light induced signals. 

\begin{figure}[tbp]
\centering
\includegraphics[width=0.49\textwidth]{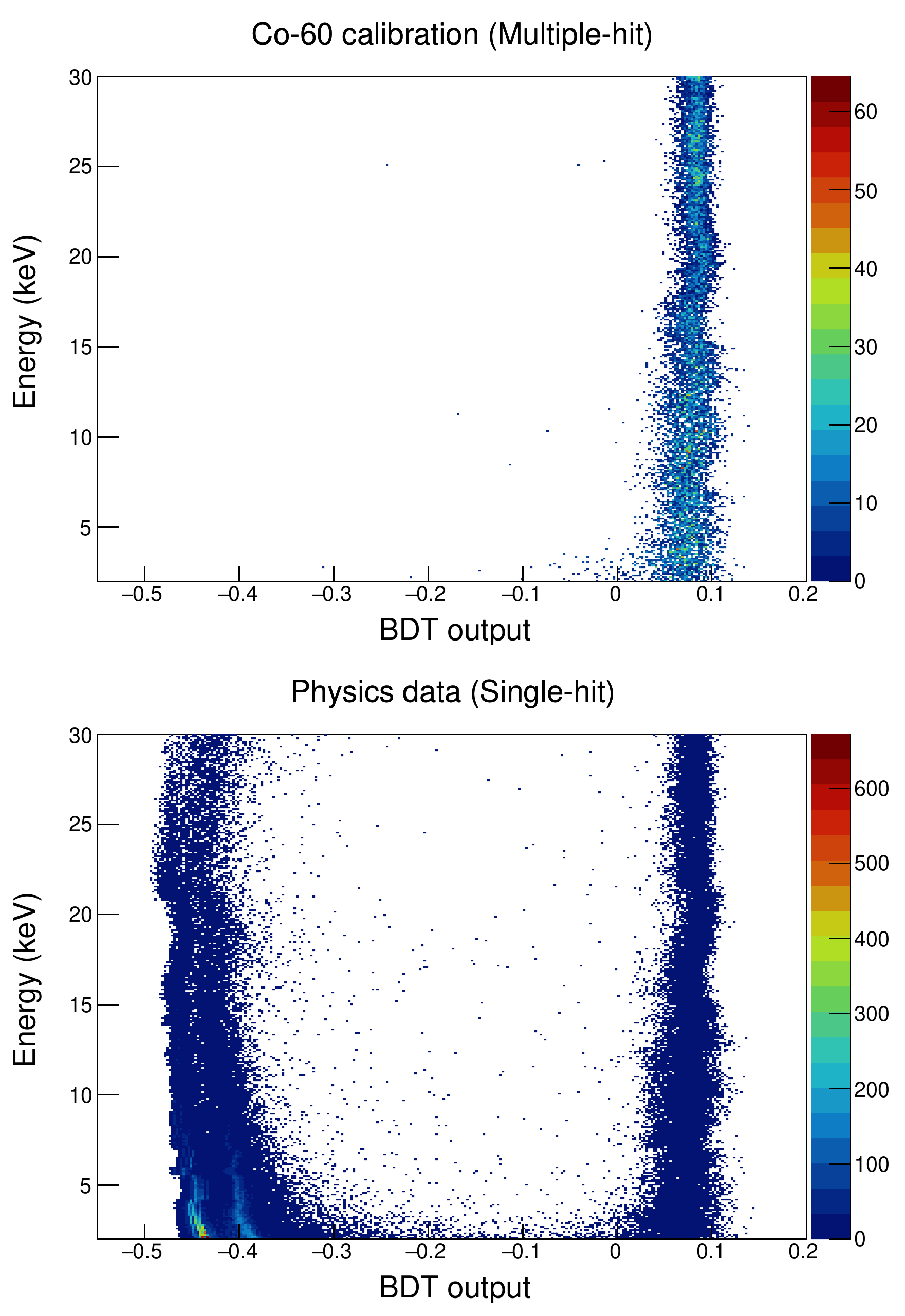} 
\caption{\label{fig_bdt}
The BDT output scores for the  $^{60}$Co calibration event sample~(top) that is used to represents scintillation signal events, and the single-hit physics data~(bottom), which represents the PMT noise-rich events with signals, are presented. The single-hit data sample shows a good separation between the PMT-induced noise (low BDT score) and the scintillation events (high BDT score) consistent with the $^{60}$Co calibration data. 
}
\end{figure}

To use these parameters, multivariate machine learning algorithms, Boosted Decision Trees~(BDTs)~\cite{BDT}, are trained to discriminate the PMT-induced noise events from radiation induced scintillation events. 
In addition to the aforementioned variables, we use the charge-weighted mean time of pulses, the total charges, and a time for 95\% charge accumulation etc. as discussed in Refs.~\cite{COSINE100:Nature,COSINE100:EFT}. 
Signal-rich multiple-hit events produced by Compton scattering of $\gamma$-rays from a $^{60}$Co calibration source, weighted to match the energy spectrum of the expected background, were used to train the scintillation signal while single-hit events from the axion search data were used for the training of the PMT-induced noises and signals. 
The BDT output for the $^{60}$Co calibration data and the single-hit physics data are shown in Fig.~\ref{fig_bdt}. This figure shows good discrimination of the PMT noises and the scintillating signals from the axion search data for energies above 2\,keV. 

\begin{figure}[!htb]
\begin{center}
\includegraphics[width=0.9\columnwidth]{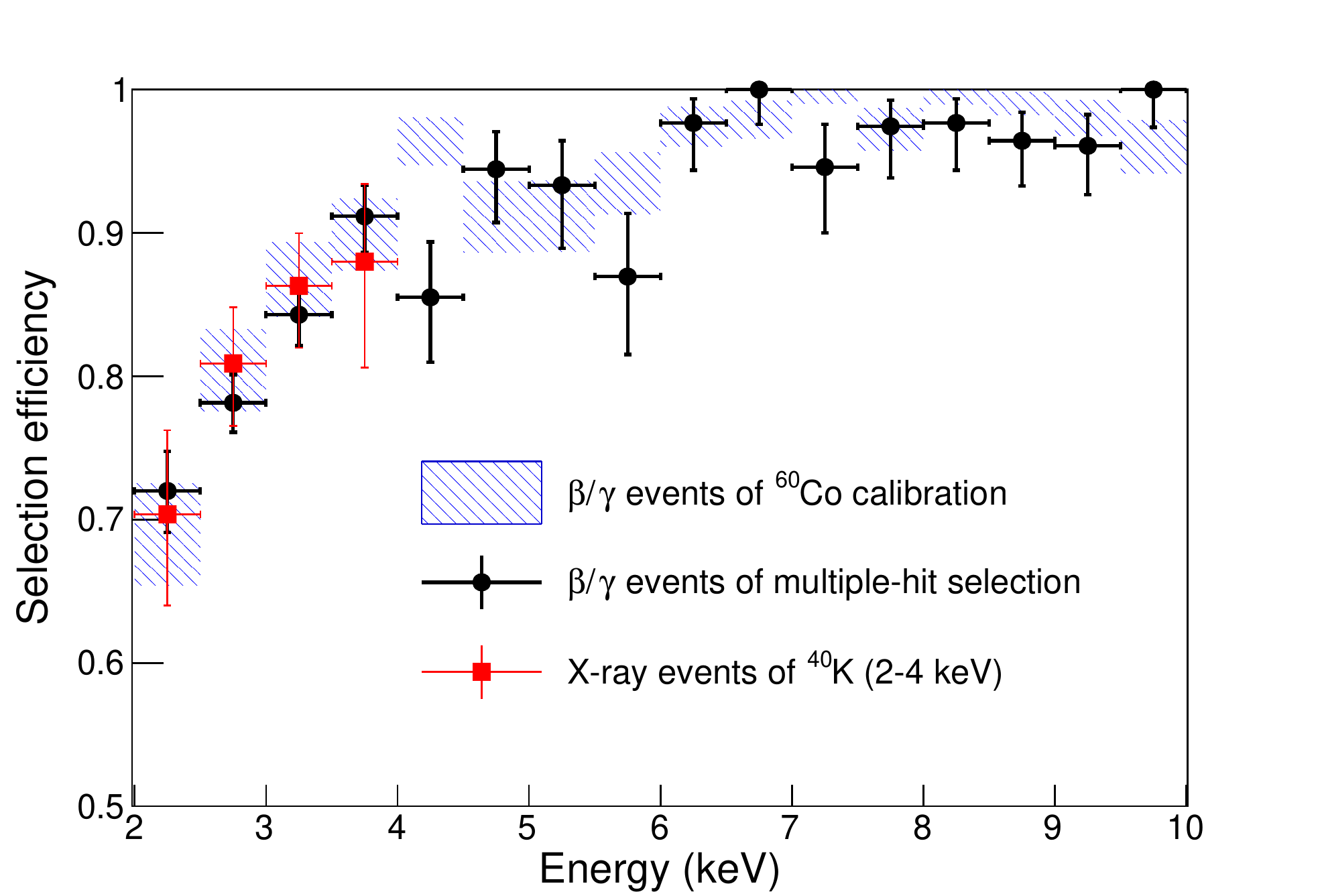}
\caption{
		(Color online)	Event selection efficiencies determined from the $^{60}$Co calibration data~(the blue statistical error band of 68\% CL), multiple-hit events of the axion search data~(black circle), and internal $^{40}$K coincident events~(red square) are presented. 
}
\label{fig_eff}
\end{center}
\end{figure}

The efficiencies of the selection criteria are determined from the $^{60}$Co source calibration data and are consistent with those measured using the multiple-hit events in the physics data that contain 3.2\,keV $^{40}$K peaks as shown in Fig.~\ref{fig_eff}. Approximately 70\,\% selection efficiency at 2\,keV is obtained for the low-threshold six crystals as one can see in Fig.~\ref{fig_eff}.

\subsection{Data analysis}
\begin{figure}[!htb]
  \begin{center}
      \includegraphics[width=0.9\columnwidth]{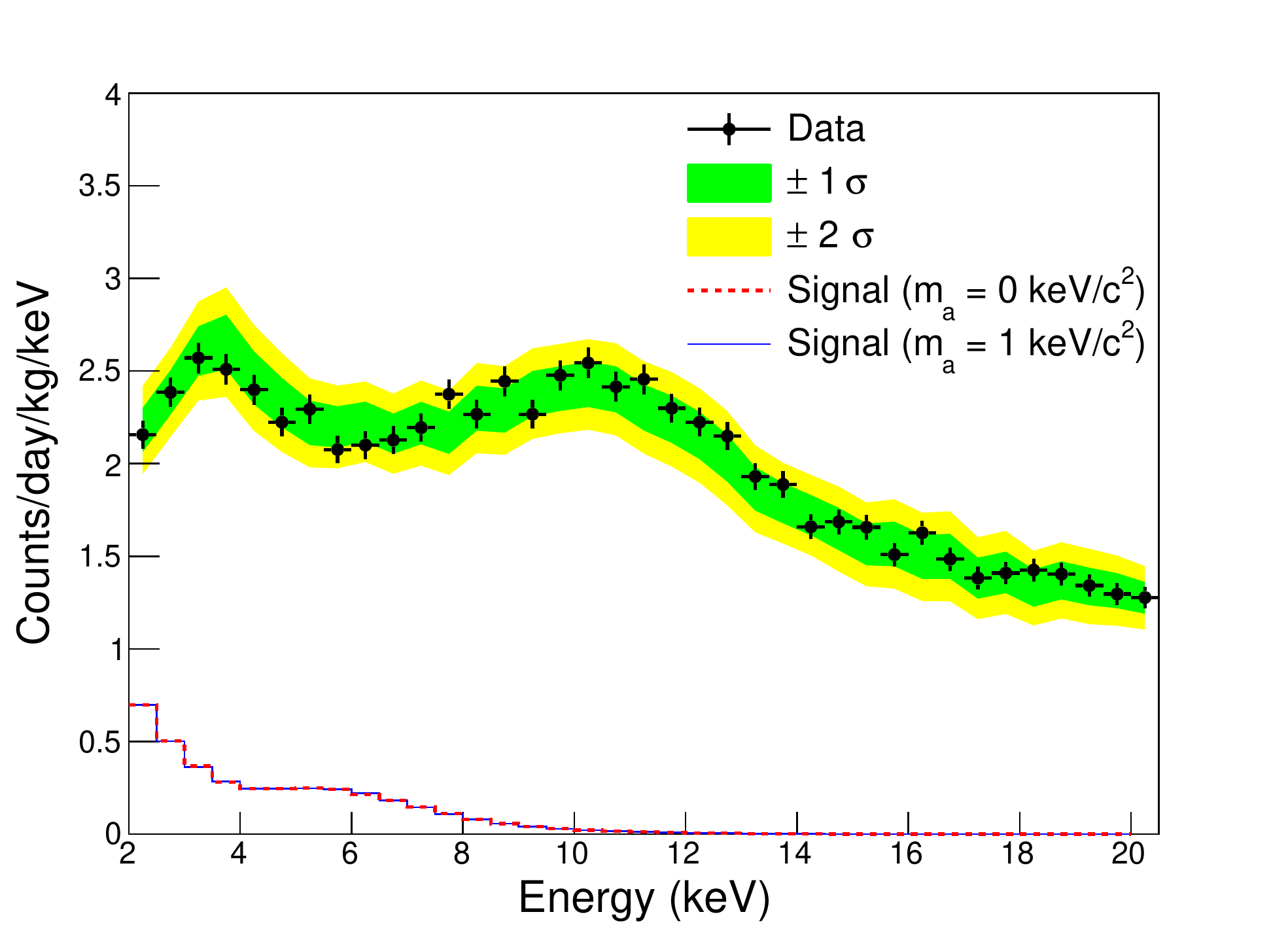} 
  \end{center}
  \caption{
  (Color online) Energy spectrum of the data with applied efficiency~(points) is compared with the predicted background spectrum for crystal-7, with 1\,$\sigma$ and 2\,$\sigma$ uncertainty bands.  The simulated axion energy spectra for $m_a$ of 0\,keV/c$^2$~(dotted red line) and 1\,keV/c$^2$~(solid blue line) for $g_{ae}$ = 1$\times10^{-10}$ are overlaid for comparison. 
	 }
  \label{eff}
\end{figure}

After application of the event selection criteria, the predominant backgrounds in the solar axion search region of interest~(ROI) are $\gamma$ and $\beta$ radiation events produced by radioactive contaminants internal to the crystals or on their surfaces, external detector components, and cosmogenic activation. These backgrounds are modeled using the Geant4~\cite{Agostinelli:2002hh} based detector simulation described in Ref.~\cite{cosinebg}. Several sources of uncertainties in the background model are included in this analysis. The largest uncertainties are those associated with the efficiency, which include statistical errors in the efficiency determination with the $^{60}$Co calibration and systematic errors derived from the independent cross-checks (see Fig.~\ref{fig_eff}). Uncertainties in the energy resolution and nonlinear responses of the NaI(Tl) crystals~\cite{nonprop}, as well as $^{210}$Pb modeling~\cite{cosinebg} are accounted as systematic uncertainties.  
Figure~\ref{eff} shows the comparison of the measured energy spectrum and the estimated one using the simulated events with their associated uncertainties.  

In the simulation program, solar axion interactions are generated using the solar axion flux and cross section discussed in section~\ref{intro}. 
Simulated events are then analyzed using the same event selection criteria as those applied to data. 
In Fig.~\ref{eff}, the simulated solar axion energy spectra for $m_a$ = 0\,keV/c$^2$ and 1\,keV/c$^2$, and for $g_{ae}$ = 1$\times$ $10^{-10}$ are overlaid on the measured background spectrum of crystal-7. No difference between  the two spectra is evident. We, therefore, use only two axion masses to describe axion signals between 0\,keV/c$^2$ and 1\,keV/c$^2$.

To estimate the solar axion signal, a binned maximum likelihood fit to the measured energy spectrum is applied, defined as,
\begin{equation}
		\mathcal{L} = \prod^{N_{ch}}_i\prod^{N_{bin}}_j \frac{\mu^{n_{ij}}_{ij}e^{-\mu_{ij}}}{n_{ij}!}\prod^{N_{bkg}}_ke^{-\frac{(x_k-\alpha_k)^2}{2\sigma_{x_k}^2}}\prod^{N_{syst}}_le^{-\frac{y_l^2}{2\sigma_{y_l}^2}},
\label{equ:llh}
\end{equation}
where $N_{ch}$ is the number of crystals,
$N_{bin}$ is the number energy bins,
$N_{bkg}$ is the number of background components,
$N_{syst}$ is the number of systematic nuisance parameters~\cite{Sinervo:2003wm},
$n_{ij}$ is the number of observed counts and $\mu_{ij}$ is the total model expectation by summing all
$N_{bkg}$ background components and a solar axion signal component after application of a shape change due to $N_{syst}$ systematic effects. In the first product of Gaussians, $x_k$ is the amount of the $k^{th}$ background component, $\alpha_k$ is the mean of $x_k$ and $\sigma_{x_k}$ is its uncertainty. In the second product of Gaussians $y_l$ is the $l^{th}$ systematic parameter and $\sigma_{y_{l}}$ is its uncertainty.
All crystals are fit simultaneously. The Bayesian Analysis Toolkit~\cite{BAT} is used with probability density functions that are based on the shapes of the simulated solar axion signals and the evaluated backgrounds. As seen in the formula, a linear prior for the signal is applied. The means and uncertainties of the Gaussian priors for the various components of the backgrounds are determined from the model fitted to the data~\cite{cosinebg}. 
To avoid biasing the axion search, the fitter was first tested with pseudo-experiments. For each axion mass, we performed 1,000 simulated experiments with the expected backgrounds and no axion signals included. From this procedure, we  calculate the expected 90\% confidence level~(CL) limits. 

\begin{figure}[!htb]
  \begin{center}
    \includegraphics[width=0.9\columnwidth]{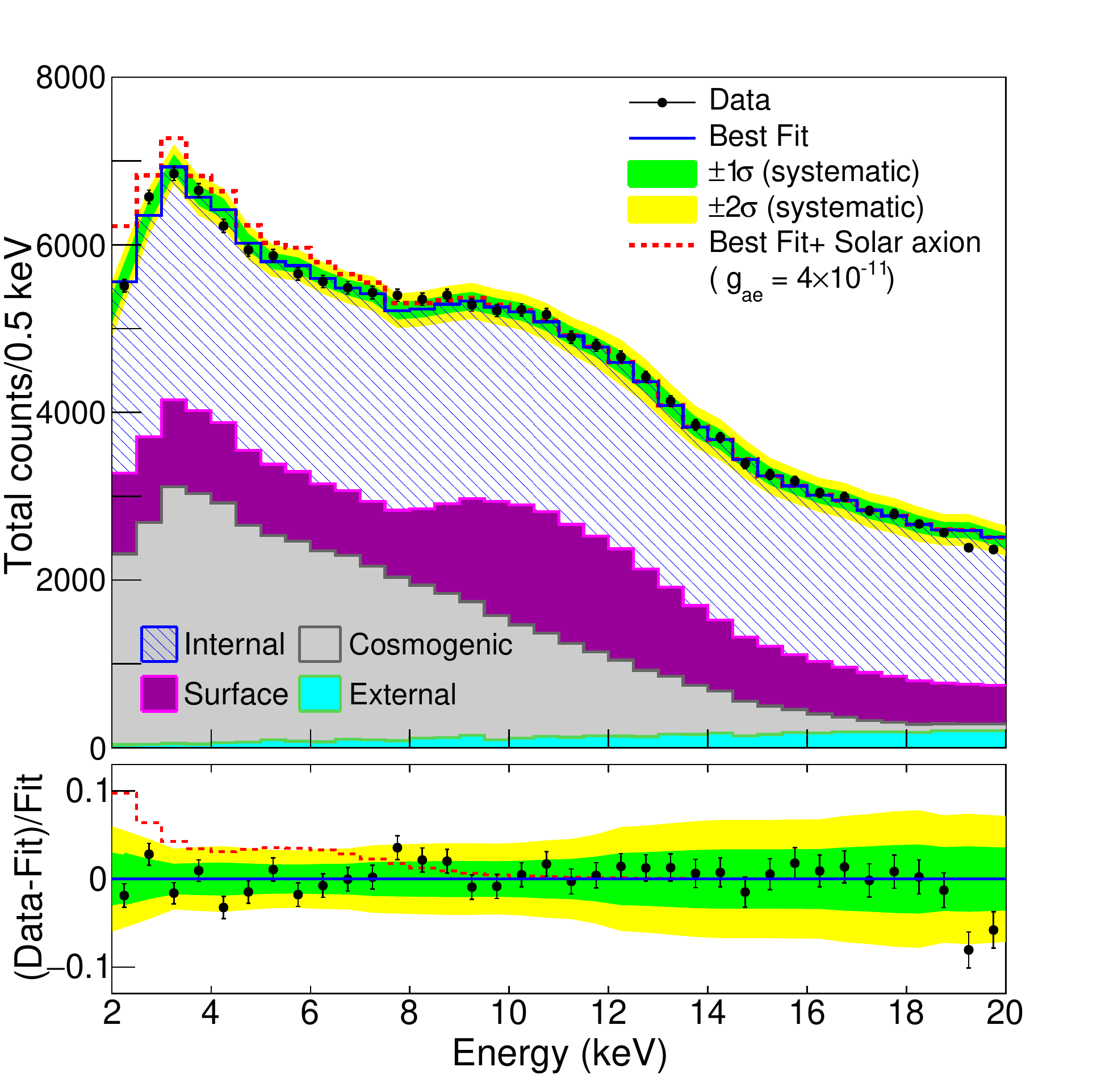}
  \end{center}
	\caption{(color online) The summed energy spectrum of the six crystals~(black points) is shown with the best fit for $m_a =0$\,keV/c$^2$~(blue solid line) overlaid with $\pm$1$\sigma$~(green) and $\pm$2$\sigma$~(yellow) shaded region of  the systematic uncertainties in the background model. For the comparison with the possible signal shape, we include a solar axion signal assuming $g_{ae}$=4~$\times$~$10^{-11}$~(red dotted line). The lower panel shows the residuals between the data and the best fit, normalized to the best fit.}
  \label{axion_spectrum}
\end{figure}

\begin{figure}[!htb]
  \begin{center}
    \includegraphics[width=0.9\columnwidth]{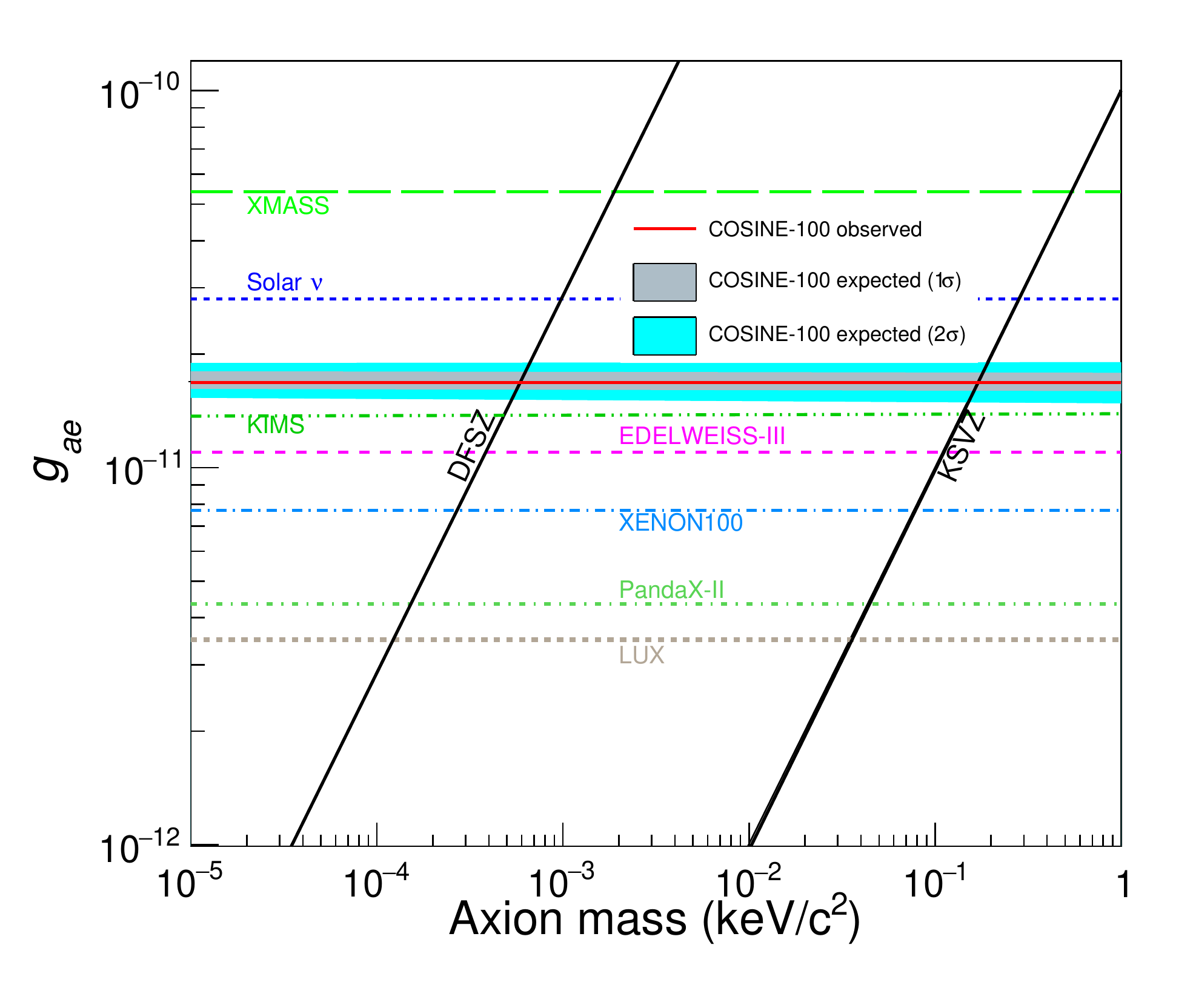}
  \end{center}
	\caption{ (Color online) The observed 90\% CL exclusion limits (red line) on the axion-electron coupling ($g_{ae}$) for the first 59.5~days data of COSINE-100 are shown together with the  68\% and 95\% probability bands for the expected 90\% CL limit assuming the background-only hypothesis. The limits are compared with the results set by XMASS~\cite{KAbe2013}, EDELWEISS-III~\cite{Armengaud:2018cuy}, KIMS~\cite{YSYoon16}, XENON100~\cite{aprile2014}, PandaX-II~\cite{Fu:2017lfc}, and LUX~\cite{Akerib:2017uem} experiments together with indirect astrophysical bounds of solar neutrino~\cite{PGondolo2009}. The inclined lines show two benchmark models of the DFSZ~($\cos^2{\beta}=1$) and KSVZ. 
 }
  \label{axion_results}
\end{figure}

Data fits are performed for two selected solar axion masses: 0\,keV/c$^2$ and 1\,keV/c$^2$. 
A maximum likelihood fit with $m_a$ = 0\,keV/c$^2$ is shown in Fig.~\ref{axion_spectrum}.
For both masses, the data fits found no excess of events that could be given by solar axion signals in our data. The posterior signal probabilities were consistent with zero in both fits. 
A limit on the axion-electron coupling of $g_{ae}$ $<$ 1.70 $\times$ $10^{-11}$ at 90\% CL for axion masses in the 0--1\,keV/c$^2$ range is set. 
The axion-electron coupling is a model-dependent coupling. In the DFSZ model, $g_{ae}$ is proportional to $\cos^2{\beta}$~\cite{Derbin:2012yk}, where $\cot{\beta}$ is the ratio of the two Higgs vacuum expectation values of the model~\cite{Raffelt:2006cw}. In the KSVZ model, it depends on $E/N$~\cite{Derbin:2012yk}, where $E$ and $N$ are the electromagnetic and color anomaly of the axial current associated with the axion field~\cite{Raffelt:2006cw}. For the KSVZ model, $E/N$ is zero as described in Ref.~\cite{EArmengaud2013} but, a much broader range of possibilities exists in axion models~\cite{Cheng:1995fd,DiLuzio:2016sbl}. For the DFSZ model, we use $\cos^2{\beta}=1$ to include most variants of the two models. These are the same choice as the recent experimental interpretations~\cite{KAbe2013,Armengaud:2018cuy,YSYoon16,aprile2014,Fu:2017lfc,Akerib:2017uem}.
QCD axions heavier than 0.59 eV/c$^2$ in the DFSZ model and 168.1 eV/c$^2$ in the KSVZ model are excluded using the parameter described above. 
Figure~\ref{axion_results} shows the observed 90\% CL limit with the $\pm1\sigma$ and $\pm2\sigma$ bands from pseudo-experiments compared with the various other experimental results. The result obtained with the COSINE-100 data is approximately five times poorer than the current world-leading limit obtained by the LUX experiment~\cite{Akerib:2017uem}. This is mainly due to a relatively small exposure, as well as a relatively large background level. A larger data-set with lower background crystals make NaI(Tl) detectors an interesting target for future solar axion searches. 

\section{Summary} 
A search for solar axions with a 59.5 day exposure of the 106 kg NaI(Tl) crystal array of the COSINE-100 dark matter search experiment has been performed. Here we apply a recent prediction for the solar axion flux that assumes that axions produce electron recoil signals in the NaI(Tl) detector through the axio-electric effects. There is no excess of events that could be attributed to solar axion interactions and this translates into an axion-electron coupling limit $g_{ae}$ $<$ 1.70 $\times$ $10^{-11}$ for axion masses in the 0--1\,keV/c$^2$ range. This excludes QCD axions heavier than 0.59\,eV/c$^2$ in the DFSZ model and 168.1\,eV/c$^2$ in the KSVZ model. 

\section*{Acknowledgments}
We thank the Korea Hydro and Nuclear Power (KHNP) Company for providing underground laboratory space at Yangyang.
This work is supported by:  the Institute for Basic Science (IBS) under project code IBS-R016-A1 and NRF-2016R1A2B3008343, Republic of Korea;
UIUC campus research board, the Alfred P. Sloan Foundation Fellowship,
NSF Grants No. PHY-1151795, PHY-1457995, DGE-1122492,
WIPAC, the Wisconsin Alumni Research Foundation, United States; 
STFC Grant ST/N000277/1 and ST/K001337/1, United Kingdom;
and Grant No. 2017/02952-0 FAPESP, CAPES Finance Code 001, Brazil.

\end{document}